\documentclass[%
 reprint,
superscriptaddress,
 amsmath,amssymb,
 aps,
pra,
]{revtex4-2}

\usepackage{graphicx}
\usepackage{dcolumn}
\usepackage{bm}
\usepackage{hyperref}

\usepackage{color}
\usepackage[dvipsnames]{xcolor}
\usepackage{ragged2e}
\usepackage[inkscapelatex=false]{svg}

\begin{document}

\preprint{APS/123-QED}

\title{Emergent Synaptic Plasticity from Tunable Dynamics of Probabilistic Bits}

\author{Sagnik Banerjee}
\email{banerj85@purdue.edu}
\affiliation{Elmore Family School of Electrical and
Computer Engineering, Purdue University, West Lafayette,
Indiana 47907, USA}

\author{Shiva T.~Konakanchi}
\affiliation{Elmore Family School of Electrical and
Computer Engineering, Purdue University, West Lafayette,
Indiana 47907, USA}

\author{Supriyo Datta}
\affiliation{Elmore Family School of Electrical and
Computer Engineering, Purdue University, West Lafayette,
Indiana 47907, USA}

\author{Pramey Upadhyaya}
\email{prameyup@purdue.edu}
\affiliation{Elmore Family School of Electrical and
Computer Engineering, Purdue University, West Lafayette,
Indiana 47907, USA}

\date{\today}

\begin{abstract}

Probabilistic (p-) computing, which leverages the stochasticity of its building blocks (p-bits) to solve a variety of computationally hard problems, has recently emerged as a promising physics-inspired hardware accelerator platform. A functionality of importance for p-computers is the ability to program--and reprogram--the interaction strength between arbitrary p-bits on-chip. In natural systems subject to random fluctuations, it is known that spatiotemporal noise can interact with the system's nonlinearities to render useful functionalities. Leveraging that principle, here we introduce a novel scheme for tunable coupling that inserts a “hidden” p-bit between each pair of computational p-bits. By modulating the fluctuation rate of the hidden p-bit relative to the synapse speed, we demonstrate both numerically and analytically that the effective interaction between the computational p-bits can be continuously tuned. Moreover, this tunability is directional, where the effective coupling from one computational p-bit to another can be made different from the reverse. This synaptic-plasticity mechanism could open new avenues for designing (re-)configurable p-computers and may inspire novel algorithms that leverage dynamic, hardware-level tuning of stochastic interactions.

\end{abstract}

\maketitle

\section{\label{sec:intro}Introduction}
 
With the increasing data- and power-hungry computing demands in today's world and the gradual decline of Moore's Law \cite{mooreslaw}, domain-specific computing platforms have become popular \cite{domainspecificAI}. To that end, several approaches that utilize physics-inspired unconventional paths to computing have generated considerable excitement lately \cite{unconventional_computing}. This shift was envisioned by Richard Feynman in his seminal 1982 paper \cite{Feynman1982}, where he proposed a quantum computer that can emulate and solve many interesting problems that involve complex probabilities and require path cancellation \cite{shor}. In the same paper, however, he also mentions a classical probabilistic (p-) computer that can deal with problems involving classical probabilities. Inspired by the same, a p-computer has recently gained attention \cite{stochastic_pbits_invert_logic_2017, pbit_qbit_book_datta, probabilistic_pbit_jan} as a natural platform to efficiently emulate inherently stochastic problems.

The fundamental building block of a p-computer is a stochastic unit called a p-bit, which produces a binary random output, $m_i$, determined probabilistically by an analog input, $I_i$. A p-computer is constructed by connecting the outputs and inputs of p-bits through synapses with coupling strengths $J_{ij}$. Such a p-computer naturally generates samples from a target probability distribution| encoded through the couplings $J_{ij}$| in $2^n-$dimensional space with $n$ p-bits. 
This feature lies at the core of efficient hardware emulation across a broad spectrum of problems—ranging from combinatorial optimization \cite{fpga2} and quantum emulation \cite{shuvro_quantum1, shuvro_quantum2, Accelerated_QMC_Shuvro} on one end, to Bayesian inference \cite{PPSL1} and machine learning \cite{fpga3} on the other. The former benefits from interpreting $m_i$ as an Ising spin. In this view, when $J_{ij} = J_{ji}$, the p-computer naturally generates samples from the Boltzmann distribution of an Ising spin system with energy $E = \frac{1}{2} \sum_{ij} J_{ij} m_i m_j$, coupled to a thermal bath \cite{boltzmann_machine_hinton_1985}. In contrast, the latter leverages the interpretation of $m_i$ as the output of a binary stochastic neuron, with the couplings $J_{ij}$ functioning as synaptic weights \cite{fullstack_pcomputing}. 

Given the promise of p-computing, a wide range of physical platforms and phenomena|such as magnetic \cite{microprocessors1, PhysRevLett.126.117202, safranski_demonstration_2021, shao_probabilistic_2023, rehm_temperature-resilient_2024, soumah2025, https://doi.org/10.1002/aelm.202001133, duffee2024, 10129319, https://doi.org/10.1002/adma.202204569, PhysRevApplied.19.024035}, optical \cite{doi:10.1126/science.adh4920, ma_quantum-limited_2025, choi_photonic_2024, horodynski_stochastic_2025, whitehead_cmos-compatible_2023}, metal-insulator transition-based \cite{metal_insulator1, metal_insulator_2}, CMOS-based \cite{pcircuits_niether_analog_nor_digital, patel2024, https://doi.org/10.1002/adfm.202307935} and digital FPGA-based \cite{fpga, fpga1, fpga2, fpga3, fpga4} systems|have recently emerged as viable candidates for implementing p-computers. A functionality of critical importance for p-computers across these platforms is the ability to tune $J_{ij}$ to generate desired probability distributions. So far, this has been largely achieved by integrating a network of p-bits with additional hardware \cite{microprocessors1, microprocessors2, linearcoupling1_series, linearcoupling1_parallel, linearcoupling2_series, linearcoupling2_parallel, programmability_sidra} specifically designed to tune $J_{ij}$. Simpler schemes allowing for on-chip tuning without the need for integration with additional components could offer more advantages. 

Engineering tunable couplings by modulating the intrinsic timescales of constituent elements have emerged as one such attractive mechanism across a broad range of unconventional computing approaches. For example, such approaches include tuning qubit frequencies to control interactions in quantum systems \cite{tunablecoupler_quantum}, modulating spiking rates in neuromorphic architectures \cite{stdp}, and adjusting oscillator frequencies in Ising spin implementations based on nonlinear oscillators \cite{tunable_coupling_oscillator}. However, to the best of our knowledge, the question of whether tuning an intrinsic timescale of p-bits can enable programmable coupling between them remains largely unexplored.

\begin{figure*}[t]
    \centering \includegraphics[width=\textwidth]{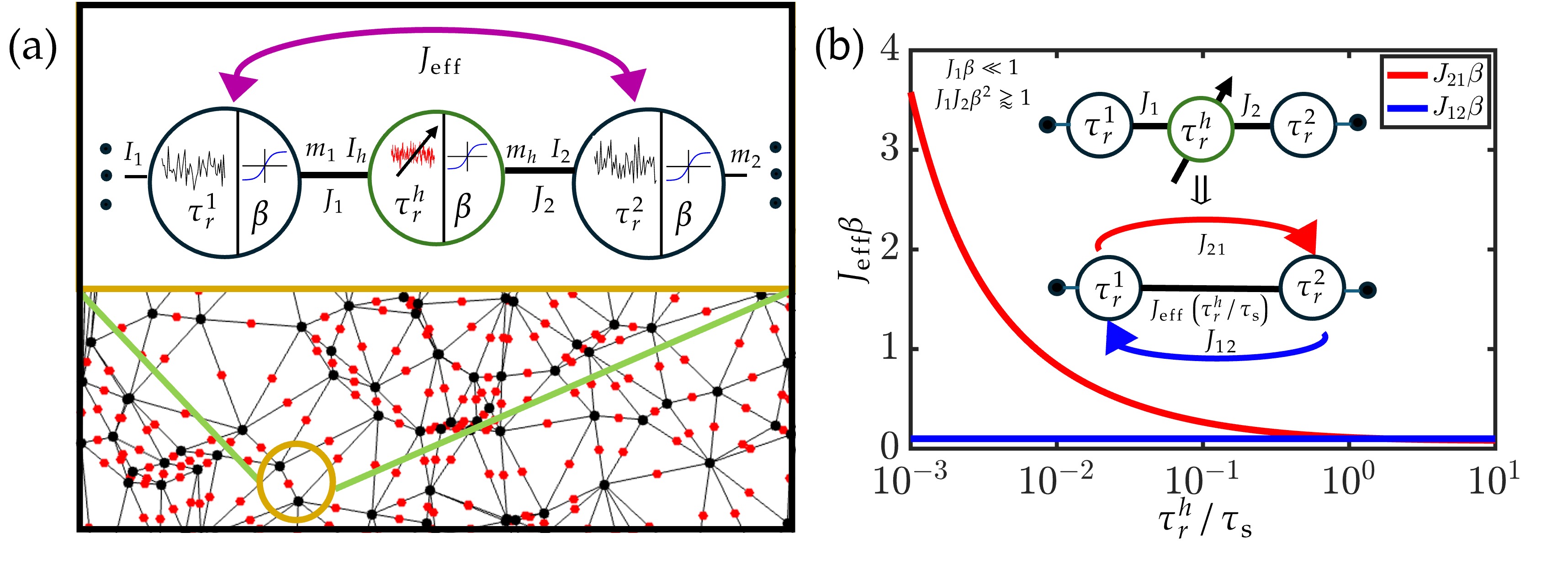}
    \caption{(a)
    Schematic representation of an interconnected p-bit network illustrating hidden p-bits (colored red) introduced between pairs of computational p-bits (colored black). A typical p-bit fluctuates randomly in time between +1 and -1. Each p-bit in the network receives continuous analog input $I_i$ and produces a binary output $m_i$.  The dynamics of each p-bit are characterized by two parameters: a correlation time $\tau_r^{i}$ and a $(k_B T)^{-1}$ like parameter $\beta$, that scales the non-linear activation function $(\operatorname{tanh})$. The computational p-bits are characterized by correlation times $\tau_r^{1}$ and $\tau_r^2$, while the hidden p-bit exhibits tunable stochasticity with adjustable correlation time $\tau_r^h$. The strength of synaptic coupling between $m_{1{(h)}}$ and $I_{h(2)}$ is labeled $J_{1(2)}$. All synapses are associated with a response delay $\tau_{\mathrm{s}}$. The effective coupling between $m_1$ and $m_2$ is depicted by $J_{\mathrm{eff}}$. (b) Programmability of $J_{\mathrm{eff}}$ between the computational p-bits in an effective model by adjusting the correlation time of the hidden p-bit relative to the synaptic delay. In the hidden p-bit model, we set $J_1 \beta \ll 1$ and $J_1 J_2 \beta^2 \gtrapprox 1$, thereby illustrating a directional asymmetry in $J_{\mathrm{eff}}$. This results in tunability of the interaction $J_{21}$ (red) but not for $J_{12}$ (blue).}
    \label{fig:1} 
\end{figure*}

In this work, we propose and theoretically analyze a scheme to engineer effective couplings between p-bits by exploiting the tunability of their intrinsic fluctuation timescales. In natural systems influenced by random fluctuations, spatiotemporal noise injection is believed to interact with nonlinear dynamics of the system to give rise to counterintuitive regular behavior \cite{spatiotemporal}. Motivated by recent developments in p-bit realizations driven by noise sources with controllable temporal correlations (see \cite{Mn3Sn_Shiva}), we design a network in which the effective coupling between any two p-bits is tuned by inserting a third p-bit, whose noise source is engineered to have faster temporal correlations than the communication time between the p-bits. We note that this regime stands in contrast to the conventional assumption in p-computing of slow p-bits and fast synapses \cite{stochastic_pbits_invert_logic_2017, pbits_PSL_2019, magnet_design_orchi_2019, PPSL1, PPSL2}. To access this unexplored domain, we extend existing behavioral models of interacting p-bits to accommodate arbitrary timescale hierarchies. Given that p-bits serve as thermally fluctuating analogs of Ising spins or binary stochastic neurons, our framework offers physical insight into how spatiotemporal fluctuation structure can mediate programmable spin-spin or interneuron interactions. On the application front, this scheme suggests a possible route to achieving synaptic programmability using native p-bit dynamics, potentially aiding scalable, on-chip implementations of probabilistic computing architectures.

\section{Main Results}
 \label{sec: main_results}
The inherent stochasticity of the p-bit arises from a noise source, which in any physical realization possesses a finite correlation time $\tau_r$. For example, in magnetic p-bits this time corresponds to the autocorrelation time of order parameter fluctuations in low-barrier nanomagnets \cite{subnanosecond_Jan}. Likewise, the communication between p-bits is mediated by synapses which have their own response time $\tau_s$. In electronic implementations, this time for example corresponds to the RC time constant of the resistive-capacitive interconnects \cite{stochastic_pbits_invert_logic_2017, pbits_PSL_2019}. 

The central idea of this work is to build a p-bit network, shown in Fig.~\ref{fig:1}(a), where slow p-bits with $\tau_r > \tau_{\mathrm{s}}$ are linked by p-bits whose fluctuation timescale $\tau_r$ can be externally tuned—either slower or faster than the fixed synaptic timescale $\tau_{\mathrm{s}}$. Different mechanisms can be used to tune $\tau_r$ depending on the physical realization. For instance, for magnetic p-bits, the rate of order parameter fluctuations can be controlled by applying a spin-orbit torque to the low-barrier magnet \cite{Mn3Sn_Shiva}. The outputs of the slow p-bits encode the probability distribution of interest and are therefore referred to as \textit{computational} p-bits. In contrast, the p-bits with tunable correlation times do not encode information themselves; rather, as we will show, they serve to modulate the effective interaction strength between computational p-bits. These are referred to here as \textit{hidden} p-bits. 

The inset in Fig.~\ref{fig:1}(a) zooms in on the basic building block of this network showing two computational p-bits, (p-bit 1 and p-bit 2) with correlation timescales $\tau_r^1$ and $\tau_r^2$ connected via a hidden p-bit with tunable $\tau_r^h$. The first and second computational p-bits are connected to the hidden p-bit with synapses of \textit{fixed} strengths $J_1$ and $J_2$, respectively. 
The main results for this proposed p-bit network can be summarized as follows.

First, we demonstrate that the basic motif|two computational p-bits connected via a hidden p-bit|can be effectively reduced to a pair of computational p-bits coupled through an interaction $J_{\rm eff}$, which can be \textit{tuned} by adjusting the ratio $\tau_r^h/\tau_{\mathrm{s}}$.
 
Second, the effective coupling mediated by the hidden p-bit can be directionally tuned, i.e., from p-bit 1 to 2 ($J_{21})$ or from 2 to 1 ($J_{12}$), by appropriately choosing the fixed dimensionless constants $J_1\beta$ and $J_2\beta$. The case for adding tunability to $J_{21}$ is shown in Fig.~\ref{fig:1}(b), where we make $J_1\beta \ll 1$ and $J_1J_2\beta^2 \gtrapprox 1$. When the hidden p-bit's correlation time is larger than the synapse ($\tau_r^h>\tau_{\mathrm{s}}$), both $J_{21}$ and $J_{12}$ converge to the weaker coupling $J_1$. In contrast, as the hidden p-bit is made faster (i.e., $\tau_r^h< \tau_{\mathrm{s}}$), deviating from the fast-synapse regime, $J_{12}$ remains equal to the weaker coupling $J_1$, but $J_{21}$ increases, becoming stronger than $J_{12}$ and $J_1$. 
 
Third, we derive analytical expressions capturing the observed dependence of the effective coupling $J_{\rm eff}$ on the ratio $\tau_r^h/\tau_{\mathrm{s}}$. Specifically, for $J_1\beta \ll 1$ and $J_1J_2\beta^2 \gtrapprox 1$, this dependence can be written as,
\begin{equation}
\begin{aligned}
J_{21} &=
\begin{cases}
\displaystyle \frac{1}{\beta} \tanh^{-1} \left[ \operatorname{erf} \left( \frac{J_1 \beta}{\sqrt{2}} \sqrt{1 + \frac{\tau_{\mathrm{s}}}{\tau_r^h}} \right) \right],~ \tau_r^h / \tau_{\mathrm{s}} \lessapprox 1, \\[10pt]
J_1,~ \tau_r^h / \tau_{\mathrm{s}} \gtrapprox 1;
\end{cases} \\
J_{12} &= J_1, \quad \forall \tau_r^h / \tau_{\mathrm{s}}.
\label{Jeff_eqn_approx}
\end{aligned}
\end{equation}
Additionally, we note that a network can also be configured to exhibit tunability in the reverse direction—i.e., where $J_{12}$ is tunable but $J_{21}$ remains fixed. This is achieved by simply swapping indices ``1" and ``2" in the earlier configuration--- specifically by choosing $J_2 \beta \ll 1$ while maintaining $J_1 J_2 \beta^2 \gtrapprox 1$. Hereafter, unless otherwise stated, we focus on the configuration where $J_1 \beta \ll 1$ and $J_1 J_2 \beta^2 \gtrapprox 1$, resulting in tunability of $J_{21}$ but not $J_{12}$.

The remainder of the paper substantiates these main results and is organized as follows. Section~\ref{sec: Model} introduces a behavioral model designed to simulate the coupled p-bit dynamics in the proposed regime. In Section~\ref{sec: synapticprogrammability}, we apply this model to conduct both numerical and analytical studies of the effective coupling between computational p-bits, highlighting its dependence on key dimensionless parameters. Finally, Section~\ref{sec: conclusion} offers concluding remarks and outlines future directions.

 \section{Model Description}
 \label{sec: Model}
To incorporate the finite correlation and synaptic response times into the simulation of collective p-bit dynamics, we adopt the approach introduced in Refs.~\cite{PPSL1, PPSL2}. Within this approach, the dynamical behavior is described by the following coupled equations:
\begin{equation}
    \begin{aligned}
m_{i}(t) &= \operatorname{sign} \{\operatorname{tanh}\left[ \beta I_{i}(t)\right] + \eta_{i}(t)\},  \\
    \end{aligned}
    \label{steady}
\end{equation}
\begin{equation}
    \begin{aligned}
\tau_{\mathrm{s}} \frac{d I_i}{d t} + I_i = \sum_{j} J_{ij} m_j.
 \\
    \end{aligned}
    \label{eq:synapse}
\end{equation}
Here, Eq.~(\ref{steady}) describes the response of p-bit: \( I_i \), the analog input to the \( i^{\text{th}} \) p-bit, controls the probability of its binary output \( m_i \) being in the \( +1 \) or \( -1 \) state. The control is modeled using a nonlinear \( \tanh(\cdot) \) activation function, whose steepness is governed by the parameter \( \beta \), while the analog-to-digital conversion is captured by the \( \text{sign}(\cdot) \) function. The inherent stochasticity of the p-bit arises from a noise source, represented by the random variable \( \eta_i \).
We also note that Eq.~(\ref{steady}) assumes that the output of the p-bit responds instantaneously to both the input and the noise source. However, in more general settings, additional timescales---such as those associated with analog-to-binary conversion or the onset of the activation function---can render the response of \( m_i \) non-instantaneous, which we ignore. Here, we instead focus on the interplay of finite correlation time of the noise source and the synapse response time. 

Eq.~(\ref{eq:synapse}) describes how p-bit outputs are transmitted via synapse. In particular, the synapse scales the output of the driving p-bit by the synaptic strength $J_{ij}$ and transmits it to the connected p-bit with a characteristic delay time $\tau_s$. This equation mimics synaptic filtering in biological networks \cite{neuroscience1} and is mostly realized in p-circuits via resistive-capacitive networks \cite{stochastic_pbits_invert_logic_2017, pbits_PSL_2019}.

Finally, to incorporate finite noise correlation time, $\eta_i(t)$ in Eq.~(\ref{steady}) is modeled as a random telegraph noise (RTN) process with an autocorrelation time $\tau_r^i$. At each time step, $\eta_i$ in the model is thus simulated by first choosing a new binary random number $r_{\rm flip,i}$ as per the following scheme.
 \begin{equation}
r_{\text{flip,i}}(t + dt) = \text{sign} \left( \exp \left( -\frac{dt}{\tau_r^i} \right) - \mathcal{U}[0,1] \right),
\label{eq: flipeqn}
\end{equation}
\noindent where $\mathcal{U} [0, 1]$ is a uniformly sampled random number between $0$ and $1$. The flip probability $p_{\text{flip,i}} = 1 - \operatorname{exp}\left({-dt/\tau_r^i}\right)$ is then approximated as a Poisson process with rate $1/\tau_r^i$.
\begin{equation}
\eta_i(t + dt) = 
\begin{cases}
\eta_i(t) & \text{if } r_{\text{flip},i} = +1 \quad (\text{no flip}), \\
\mathcal{U}[-1, 1] & \text{if } r_{\text{flip},i} = -1 \quad (\text{flip}). 
\label{eq:RTNeqn}
\end{cases}
\end{equation}
If $r_{\text{flip}, i}=-1$, a new random number is assigned to $\eta_i(t + dt)$ by sampling from the uniform distribution $\mathcal{U}[-1, 1]$. However, for $r_{\text{flip}, i}=+1$, our model retains the previously computed $\eta$ for the next time step. Such an update rule ensures $\eta_i(t)$ retains memory over timescales $\sim \tau_r^i$ and mimics RTN behavior for $\eta_i(t)$ with finite autocorrelation time. This $\eta_i(t)$ is then fed into Eq.~(\ref{steady}), which is solved numerically in combination with Eq.~(\ref{eq:synapse}) to simulate the dynamics of the p-bit network.

This model offers several key advantages for our purpose. First, as demonstrated in Ref.~\cite{PPSL1}, a similar behavioral model has successfully reproduced results from full time-domain device-level simulations of stochastic magnetic tunnel junction (s-MTJ)-based p-bits. Given that behavioral models are significantly more computationally efficient and compact, they enable a thorough exploration of the phase space, which is critical for understanding the behavior of our proposed networks. Notably, since most prior work in p-computing has focused on the so-called fast-synapse regime (i.e., $\tau_r > \tau_s$), existing models typically evolve synaptic states using update rules derived by analytically solving Eq.~(\ref{eq:synapse}) in the fast synapse regime \cite{PPSL1}. However, this analytical solution is not valid in the fast p-bit regime (i.e., $\tau_r < \tau_s$). Since here we allow for hidden p-bits to go in the fast p-bit regime (i.e., $\tau_r < \tau_s$) we cannot use the update rules derived in previous studies. To accommodate this, we solve the coupled differential equations numerically, allowing for arbitrary hierarchies in timescales. Second, with suitable choices of $\tau_r$, $\tau_s$, $J_{ij}$, and $\beta$, the model retains broad applicability across different physical implementations of p-circuits, enabling system-agnostic investigations.

 \section{Synaptic Programmability}
 \label{sec: synapticprogrammability}

\begin{figure*}[t]
    \centering \includegraphics[width=\textwidth]{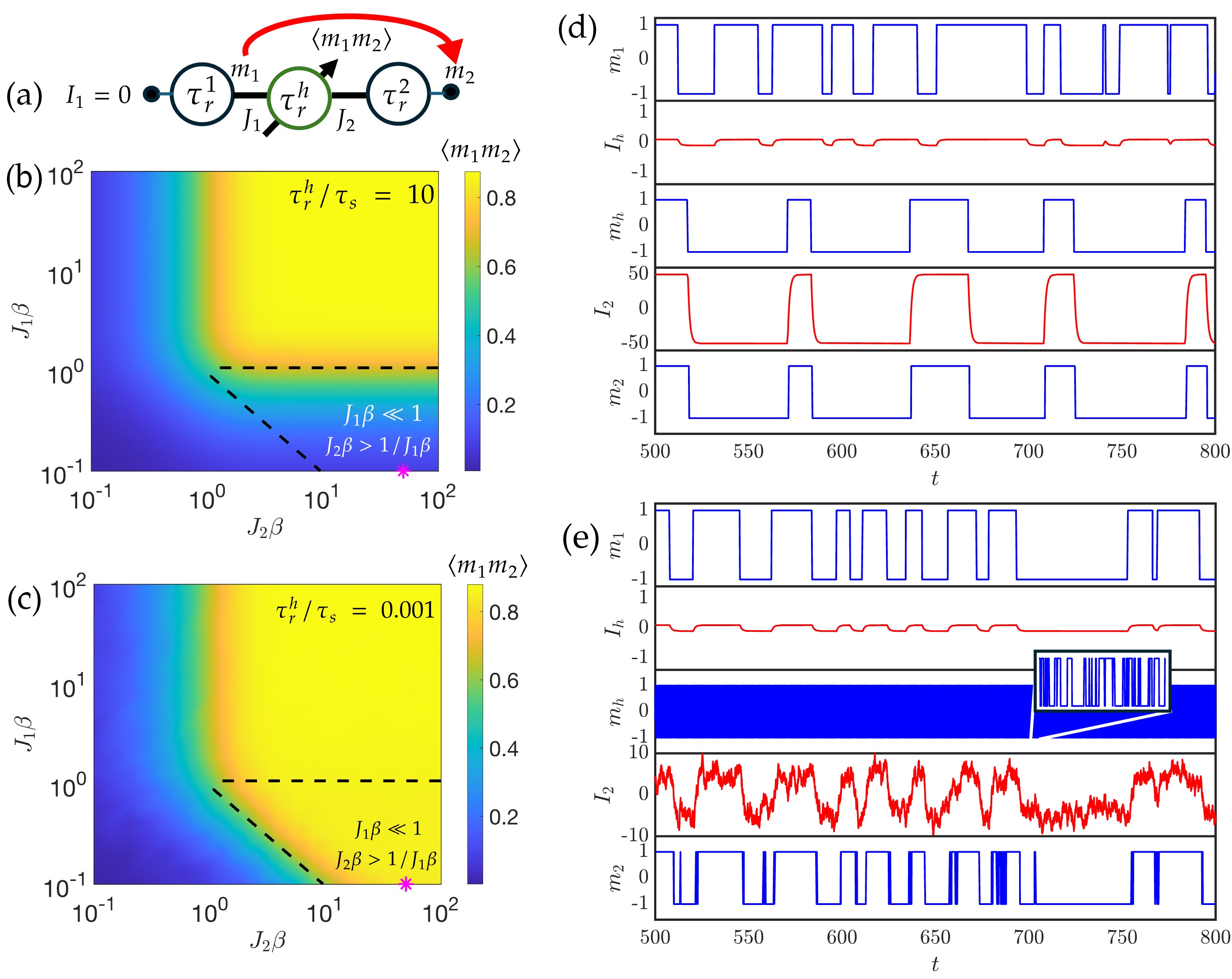}
    \caption{(a) Schematic representation of the hidden p-bit model under investigation. Here, the input of p-bit 1 has been pinned to $zero$, resulting in its output $m_1$ fluctuating stochastically with a correlation time governed by $\tau_r^1$. The time-averaged correlator $\langle m_1 m_2 \rangle$ between the outputs of p-bits 1 and 2 quantifies the effective coupling between them mediated by the hidden p-bit. Phase diagram illustrating the dependence of $\langle m_1 m_2 \rangle$ on $J_1 \beta$ and $J_2 \beta$ for two distinct regimes: (b) $\tau_r^h/\tau_{\mathrm{s}} = 10$ and (c) $\tau_r^h/\tau_{\mathrm{s}} = 0.001$. $\langle m_1 m_2 \rangle$ demonstrates tunability from (b) to (c) only in the region where $J_1 \beta \ll 1$ and $J_2 \beta > 1/J_1 \beta$ (enclosed by the dashed line). However, this asymmetry is directional, so no correlation change is observed for $J_1 \beta > 1/J_2 \beta$ and $J_2 \beta \ll 1$. For specific choices of parameters $J_1 \beta = 0.1$ and $J_2 \beta = 50$ in the phase diagram (denoted by asterisk), the time evolutions in the inputs and outputs of the p-bits highlight the distinct behaviors in $\langle m_1 m_2 \rangle$: (d) weak correlations for $\tau_r^h/\tau_{\mathrm{s}} = 10$ but (e) strong correlations for $\tau_r^h/\tau_{\mathrm{s}} = 0.001$. A zoomed inset constructed in (e) represents an ultra-fast signature observed in $m_h$, unlike in (d). The correlation time for the computational p-bits has been set to $\tau_r = 10$ and the synaptic response time is assumed to be $\tau_{\mathrm{s}} = 1$ everywhere.}
    \label{fig:2}
\end{figure*}

\textbf{A. Hidden p-bit model:} Having discussed the numerical model, we next apply it to study the collective dynamics of the basic motif introduced in Sec.~\ref{sec: main_results}|two computational p-bits connected via a hidden p-bit (which we refer to here as the hidden p-bit model). Specifically, to probe the central quantity of interest in this work|the effective coupling between computational p-bits|we focus on the case where one computational p-bit is allowed to fluctuate randomly between $+1$ and $-1$, and we monitor how faithfully the other computational p-bit tracks its state. Without loss of generality, we designate p-bit 1 as the random fluctuator by pinning its input to $I_1 = 0$, and quantify the effective coupling by calculating the time-averaged correlator between the outputs of the two computational p-bits [see Fig.~\ref{fig:2}(a)]:
\begin{equation}
     \langle m_1 m_2 \rangle = \frac{1}{N}  \sum_{k=1}^N m_1(k) \cdot m_2(k),
 \end{equation}
as obtained by numerically evolving the p-bits according to Eqs.~\eqref{steady}--~\eqref{eq:RTNeqn}. Here $\langle.\rangle$ is the time-average taken over $N$ simulation steps. The coupled dynamics generically depends on the following dimensionless parameters of the hidden p-bit model: $J_1\beta$, $J_2\beta$, $\tau_r^1/\tau_{\mathrm{s}}$, $\tau_r^2/\tau_{\mathrm{s}}$, and $\tau_r^h/\tau_{\mathrm{s}}$. In our proposed scheme, the computational p-bits $\tau_r$ are fixed and designed to be much slower than the synapse response time; accordingly, we set $\tau_r^1/\tau_{\mathrm{s}}=\tau_r^2/\tau_{\mathrm{s}} \gg 1$ throughout and study the dynamics as a function of $J_1\beta$, $J_2\beta$ and $\tau_r^h/\tau_{\mathrm{s}}$.

Figs.~\ref{fig:2}(b) and \ref{fig:2}(c) show the results of $\langle m_1 m_2 \rangle$ as a function of $J_1\beta$ and $J_2\beta$ for two limiting cases: slow hidden p-bits ($\tau_r^h/\tau_{\mathrm{s}} = 10$) and fast hidden p-bits ($\tau_r^h/\tau_{\mathrm{s}} = 0.001$), respectively. It is observed that strong correlations consistently emerge when \( J_{1(2)}\beta > 1 \), while weak correlations prevail in the regime \( J_{1(2)}\beta < 1 \), largely independent of the hidden p-bit's correlation time. However, the phase space enclosed by \( J_1\beta \ll 1 \) and \( J_2 \beta > 1/J_1 \beta \) exhibits a pronounced sensitivity to the hidden p-bit's correlation time. Interestingly, this sensitivity is not present in the opposite limit—enclosed by \( J_2\beta \ll 1 \) and \( J_1 \beta > 1/J_2 \beta \). The emergence of strong and weak correlations in the \( J_{1(2)}\beta > 1 \) and \( J_{1(2)}\beta < 1 \) regimes, respectively, align with the expectations based on the coupling strengths. However, the sensitivity of $\langle m_1 m_2 \rangle$ to the hidden p-bit's correlation time in the $J_1 \beta \ll 1$ and $J_1 J_2 \beta^2 \gtrapprox 1$ case is not immediately obvious. Additionally, the insensitivity of \( \langle m_1 m_2 \rangle \) in the \( J_2\beta \ll 1 \) and $J_1 J_2 \beta^2 \gtrapprox 1$ regime remains nontrivial and highlights the directional nature of the effective coupling mediated by the hidden p-bit. This also demonstrates that, with appropriate choices of $J_1\beta$ and $J_2\beta$, the effective coupling between computational p-bits can be tuned by adjusting the flipping rate of the hidden p-bit.


To gain more insight into the tunable coupling for $J_1 \beta \ll 1$ and $J_1J_2 \beta^2 \gtrapprox 1$, in Figs.~\ref{fig:2}(d) and \ref{fig:2}(e) we show the time evolution of the p-bits inputs and outputs for a specific choice of parameters $J_1 \beta = 0.1$ and $J_2 \beta = 50$ within this regime, for the $\tau_r^h/\tau_{\mathrm{s}} = 10$ and $\tau_r^h/\tau_{\mathrm{s}}=0.001$ cases, respectively. For the case when hidden p-bit updates slower than the synapse's response [Fig.~\ref{fig:2}(d)], all the p-bits in the motif are in the conventional fast-synapse regime. Consequently, the synapse is able to communicate the \textit{instantaneous} output state of the p-bit driving the synapse to the input of the p-bit it connects to, with the transmitted signal scaled by the strength of the coupling between the p-bits. Depending on the coupling strength|whether $J\beta \ll 1$ or $J\beta \gg 1$| the two p-bits will exhibit weak or strong correlation, respectively. This is clearly seen in Fig.~\ref{fig:2}(d), where $J_1 \beta \ll 1$ results in weak tracking between $m_1$ and $m_h$, while the strong coupling $J_2 \beta \gg 1$  leads to strong correlation between $m_h$ and $m_2$. As a result, in this case $\langle m_1m_2\rangle \sim \langle m_1m_h\rangle \ll 1$.

In contrast, when the hidden p-bit updates faster than the synapse's response time [Fig.~\ref{fig:2}(e)], the above picture can no longer be used. While the instantaneous output of the first slow computational p-bit can still be scaled by $J_1$ and transferred to the hidden p-bit's input [see $I_h$ in Fig.~\ref{fig:2}(e)], the hidden p-bit's rapid updates prevent the synapse from communicating the instantaneous value of $m_h$ to the second computational p-bit. Instead, the second computational p-bit receives the hidden p-bit’s \textit{filtered} output, shaped by the synapse's RC-like response. Interestingly, this filtered response is an amplified (albeit noisy) version of $m_1$ [see $I_2$ in Fig.~\ref{fig:2}(e); an analytical explanation of this filtering effect is provided in appendix~\ref{supp: 3pbitmodel}]. As per Eq.~(\ref{steady}), $m_2$ tends to follow a strong input $I_2$, and $I_2$ resembles $m_1$|this leads to an increase in $\langle m_1 m_2 \rangle$ towards unity. 
 
Another important feature of the hidden p-bit mediated tuning is its inherent asymmetry: as seen in Figs.~\ref{fig:2}(b) and (c), tuning is effective when $J_1\beta$ is small and $J_2\beta$ is large, and not the other way around. In the other limit, when $J_1\beta > 1$ the input to the hidden p-bit becomes strongly biased by $m_1$, effectively pinning the hidden p-bit’s output to $m_1$ via Eq.~(\ref{steady}). As a result, the hidden p-bit’s internal noise $\eta_h$ and its correlation time become irrelevant, resulting in the loss of the intended tunability. A striking consequence of this asymmetry is that it introduces directionality in the coupling modulation: for a fixed design with  $J_1\beta \ll 1$ and $J_1J_2 \beta^2\gtrapprox 1$, tuning the effective coupling works when $m_1$ drives $m_2$, but not in the reverse direction. 

\begin{figure*}[t]
    \centering \includegraphics[width=\textwidth]{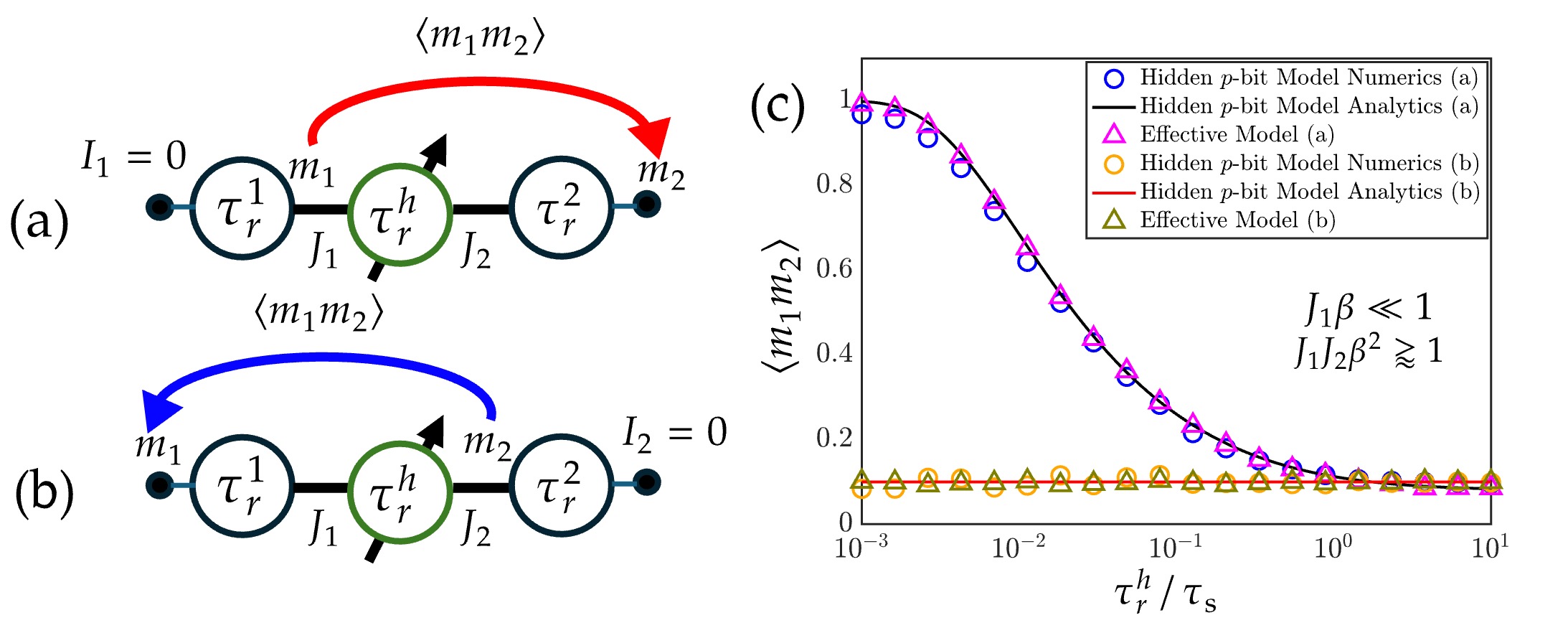}
    \caption{Directional asymmetry in the hidden p-bit model illustrated for two cases: (a) input to p-bit 1 fixed to $zero$ ($I_1=0$) enabling random fluctuations in $m_1$ and causing correlations to $m_2$ controlled via the hidden p-bit. (b) Input to p-bit 2 is pinned to $zero$, leading to random fluctuations in $m_2$ which is always uncorrelated to $m_1$. For both these cases, $J_1 \beta \ll 1$ and $J_1 J_2 \beta^2 \gtrapprox 1$. (c) Variation in $\langle m_1 m_2 \rangle$ with $\tau_r^h/\tau_{\mathrm{s}}$, demonstrating tunable correlations for case (a) and consistently negligible correlations for case (b). Numerical simulations of the hidden p-bit model with slow computational p-bits $1$ and $2$ (blue and yellow: $\tau_r/\tau_{\mathrm{s}} = 100$) agree with the analytical theory (black and red lines) derived for $\tau_r \gg \tau_{\mathrm{s}}$, in both the cases. The hidden p-bit model is further validated by comparison with an effective model (magenta and green) characterized by a synaptic coupling strength $J_{\mathrm{eff}}$.}
    \label{fig:3}
\end{figure*}

$\textit{Analytics.}$| We next proceed to develop an analytical understanding of the $\langle m_1 m_2 \rangle$ correlators for the hidden p-bit model. To this end, first note that for $I_1=0$ in Eq.~(\ref{steady}), $m_1 = \mathrm{sign}(\eta_1)$, meaning $m_1$ behaves as a binary random telegraph noise (RTN) fluctuating on the computational p-bit timescale $\tau_r^1$. Since we are interested in the limit where $\tau_r^1$ is much larger than the other timescales ($\tau_{\mathrm{s}}$, $\tau_r^h$), for calculating $\langle m_1 m_2 \rangle$ we can average over timescales where $m_1$ can be treated as a constant. Without loss of generality, we set the value of $m_1$ to $+1$. Consequently, $I_h \approx J_1$ for this subsection and evaluating $\langle m_1 m_2 \rangle$ reduces to calculating $\sim \langle m_2 \rangle$.

Second, motivated by the tunability and directionality features highlighted by the numerics, we focus on calculating $\langle m_2 \rangle$ as a function of the ratio $\tau_r^h/\tau_{\mathrm{s}}$ in the following two regimes:

\noindent (i) $J_1\beta \ll1$ and $J_1J_2\beta^2 \gtrapprox 1$ — As seen in Fig.~\ref{fig:2}, this corresponds to the regime where the correlators exhibit strong tunability with $\tau_r^h/\tau_{\mathrm{s}}$. We first consider the limit $\tau_r^h/\tau_{\mathrm{s}} \gtrapprox 1$. In this case, the synapse can transmit instantaneous outputs, and combined with the large $J_2\beta$ inherent to this regime, we have $m_2 \sim m_h$. Thus, $\langle m_2 \rangle$ can be evaluated by averaging over $m_h$. Using $I_h \approx J_1$ in Eq.~(\ref{steady}) for the hidden p-bit, we obtain, (see appendix~\ref{supp: Effective2pbitmodel}): \begin{equation} \langle m_1 m_2 \rangle \sim \tanh(\beta J_1), \quad \tau_r^h/\tau_{\mathrm{s}} \gtrapprox 1. \label{eq:mimijformula1} \end{equation}
In contrast, for $\tau_r^h/\tau_{\mathrm{s}} \lessapprox 1$, $m_2$ can no longer instantaneously follow $m_h$. Instead, it is driven by a stochastic input $I_2$, which is obtained after applying an RC-like low-pass filter to $m_h$ with a DC gain $J_2$ [see Eq.~(\ref{eq:synapse})]. In the limit of a fast hidden p-bit, the filtered $I_2$ can be approximated as a Gaussian variable with mean $\mu = J_1J_2\beta$ and standard deviation $\sigma = J_2/\sqrt{1+\tau_{\mathrm{s}}/\tau_r^h}$ (see appendix~\ref{supp: 3pbitmodel}). Moreover, since $J_1J_2\beta^2 \gtrapprox 1$, the local noise term $\eta_2$ can be neglected. Therefore, $\langle m_2 \rangle$ can be evaluated as (see appendix~\ref{supp: 3pbitmodel}):
\begin{equation}
    \langle m_1 m_2 \rangle  = \operatorname{erf} \left(\frac{J_1 \beta}{\sqrt{2}} \sqrt{1 +  \frac{\tau_{\mathrm{s}}}{\tau_r^h}}\right);~\tau_r^h/\tau_{\mathrm{s}} \lessapprox 1.
    \label{eq:mimijformula2}
\end{equation}
Here, $\operatorname{erf}(x)$ is the error function defined as $\operatorname{erf}(x) = \frac{2}{\sqrt{\pi}} \int_0^x e^{-t^2}\, dt$.

\noindent(ii) $J_2\beta \ll 1$ and $J_1 J_2\beta^2 \gtrapprox 1$| In this case, the correlators become independent of $\tau_r^h/\tau_{\mathrm{s}}$, illustrating the directionality aspect of tunability. To see this analytically, we note that using Eq.~(\ref{steady}) for the hidden p-bit, the large $J_1\beta$ characteristic of this regime enforces $m_h \sim m_1 = 1$. Crucially, this pinning of $m_h$ occurs independently of the local noise at the hidden p-bit and is thus unaffected by the ratio $\tau_r^h/\tau_{\mathrm{s}}$. Consequently, the input to the second computational p-bit becomes $I_2 \approx J_2$, yielding an average $\langle m_2 \rangle$, and therefore the correlator $\langle m_1 m_2 \rangle$ as (see appendix~\ref{supp: Effective2pbitmodel}):
\begin{equation}
    \langle m_1 m_2 \rangle  \sim \tanh(\beta J_2);~ \forall \tau_r^h/\tau_{\mathrm{s}}. 
    \label{eq:mimijformula3}
\end{equation}

Eqs.~(\ref{eq:mimijformula1}), (\ref{eq:mimijformula2}), and (\ref{eq:mimijformula3}) summarizes the main analytical results for the hidden p-bit model. To validate this analytical understanding and demonstrate the directional tunability of $\langle m_1 m_2 \rangle$ induced by tuning $\tau_r^h/\tau_{\mathrm{s}}$, we compare analytics and numerics for two cases with $J_1\beta \ll 1$ and $J_1J_2\beta^2 \gtrapprox 1$: (i) when p-bit 1 drives p-bit 2, by setting $I_1=0$ and calculating $\langle m_1 m_2 \rangle$ [schematically depicted in Fig.~\ref{fig:3}(a)], and (ii) when p-bit 2 drives p-bit 1, by setting $I_2=0$ and calculating $\langle m_1 m_2 \rangle$ [Fig.~\ref{fig:3}(b)]. As described in Sec.~\ref{sec: main_results}, (ii) is equivalent to (i) for $J_2 \beta \ll 1$ and $J_1 J_2 \beta^2 \gtrapprox 1$ and can be obtained by swapping the indices ``1" and ``2" in (i). 

As shown in Fig.~\ref{fig:3}(c), when p-bit 1 drives p-bit 2, initially correlated p-bits can be smoothly decorrelated by tuning $\tau_r^h/\tau_{\mathrm{s}}$. In contrast, when p-bit 2 drives p-bit 1, the correlators remain low and non-tunable. Lastly, the analytical results match closely with the numerical simulations, corroborating the proposed mechanisms. \\


\textbf{B. Effective model:} We next turn to quantitatively extract the strength of the effective tunable coupling $J_{\mathrm{eff}}$ mediated by hidden p-bits. To this end, we map the basic building block|two computational p-bits coupled via a flipping-rate-tunable hidden p-bit|onto an equivalent model of two computational p-bits coupled via an effective direct coupling $J_{\mathrm{eff}}$ [see the inset of Fig.~\ref{fig:1}(b); referred here as the effective model]. 

To establish this mapping, for a given set of hidden p-bit model parameters ($J_1$, $J_2$, $\beta$, $\tau_r/\tau_{\mathrm{s}}$), we determine the value of $J_{\mathrm{eff}}$ in the effective model that matches the $\langle m_1 m_2 \rangle$ correlations of the hidden p-bit model. Additionally, we perform this mapping for both cases: when p-bit 1 drives p-bit 2, denoting the effective coupling as $J_{21}$, and when p-bit 2 drives p-bit 1, denoting the effective coupling as $J_{12}$. Without loss of generality, we consider the case when $I_1 = 0$, then the probability of $m_1$ being $+1$ or $-1$ are equal, i.e., $p(m_1 = +1) = p(m_1 = -1) = 1/2$. In the limit $ \tau_r^1/\tau_{\mathrm{s}} \gg 1$, the input to p-bit $2$ is $I_2 \approx J_{\mathrm{eff}} m_1$. The correlator can be written analytically as (see appendix~\ref{supp: Effective2pbitmodel}): $\langle m_1 m_2 \rangle= \tanh(\beta J_{\rm eff})$. Equating this to the analytical correlators of the hidden p-bit model in the appropriate regimes highlighted in Eqs.~\eqref{eq:mimijformula1} --\eqref{eq:mimijformula3}, we arrive at Eq.~(\ref{Jeff_eqn_approx}) (introduced as a main result in Sec.~\ref{sec: main_results}).
In Fig.~\ref{fig:3}(c), we plot both the analytical and numerical correlators of the effective model using $J_{\mathrm{eff}}$ extracted from Eq.~(\ref{Jeff_eqn_approx}), showing good agreement with the hidden p-bit model and validating the extracted $J_{\mathrm{eff}}$. 

\section{Outlook}
\label{sec: conclusion}
In summary, we propose and theoretically demonstrate a scheme that leverages the fluctuating timescales of p-bits as a new knob for enabling on-chip programmability in p-computers. From an application perspective, a key advantage of our approach is that it uses p-bits themselves as couplers, allowing integration into existing p-circuit architectures without requiring additional hardware. For instance, our scheme could be emulated in current FPGA-based p-circuit implementations, where clocks with tunable rates can serve as the knob for adjusting p-bit correlation times to program the couplings \cite{fpga2}. In more scalable implementations involving emerging nanoscale devices, such as stochastic magnetic tunnel junction-based p-bits, spin-orbit torque-induced control of magnetic fluctuation rates can be utilized to realize our proposal \cite{Mn3Sn_Shiva}. In such systems, the flipping rate of a p-bit can be tuned by current-induced modifications of an effective magnetic energy barrier. Given the exponential dependence of the flipping rate on the barrier height, several orders of magnitude changes in flipping rates are achievable within existing platforms, allowing for significant tunability of the correlations. Antiferromagnet-based p-bits, which have the potential to reach picosecond-level flipping rates \cite{Mn3Sn_Shiva, Konakanchi2024, Banerjee2024, subnanosecond_Jan}, could be particularly promising candidates for implementing this scheme.

Moreover, the directional nature of the tuning could also be advantageous. For example, one can envision creating a p-circuit network where computational p-bits are connected by two hidden p-bits: one designed to mediate coupling from computational p-bit \( j \) to computational p-bit \( i \) (denoted as \( J_{ij} \)), and the other for mediating coupling in the reverse direction (\( J_{ji} \)). By separately tuning the hidden p-bits, it is possible to program the coupling \( J_{ij} \) while ensuring \( J_{ij} = J_{ji} \). This symmetric coupling would enable the resultant network to be used for applications requiring programmable energy functions of the form $E = \frac{1}{2} \sum_{i,j} J_{ij} m_i m_j$, such as sampling from the Boltzmann distribution~\cite{boltzmann_machine_hinton_1985} and annealing based optimization  ~\cite{intrinsic_optimization_sutton}. On the other hand, one could also program couplings with \( J_{ij} \neq J_{ji} \), enabling the design of programmable directed networks. Such networks could find applications in machine learning and Bayesian inferencing tasks~\cite{PPSL1}.

 \begin{acknowledgments}
 We would like to thank Risi Jaiswal, Kerem Camsari, Shunsuke Fukami and Joerg Appenzeller for helpful discussions. SB, STK and PU acknowledge support from the National Science Foundation (NSF) grant DMREF-2324203. STK acknowledges support from the NSF grant ECCS-2331109. SD acknowledges support from ONR-MURI Grant No.~N000142312708, OptNet: Optimization with p-Bit Networks.
 \end{acknowledgments}

 \section*{Conflict of Interest}
SD has a ﬁnancial interest in Ludwig Computing.

\appendix
\renewcommand{\thefigure}{A\arabic{figure}}
\setcounter{figure}{0}

\section{ Hidden p-bit model analytics}
\label{supp: 3pbitmodel}
In this section, we build an analytical understanding of the hidden p-bit model in the $\tau_r^h/\tau_{\mathrm{s}} \lessapprox 1$ regime, for the directionally tunable case of $J_1 \beta \ll 1$ and $J_1 J_2 \beta^2 \gtrapprox 1$, to arrive at Eq.~(\ref{eq:mimijformula2}).

To calculate $\langle m_1 m_2 \rangle$ in the limit of p-bit 1 driving p-bit 2, we set $I_1 = 0$ in Eq.~(\ref{steady}) for simplicity. $m_1 = \operatorname{sign}(\eta_1)$ then is a binary RTN with correlation timescale of $\tau_r^1$. Since we are interested in timescales $\tau_r^1 \gg \tau_{\mathrm{s}} \gtrapprox \tau_r^h$, $\langle m_1 m_2 \rangle$ can be calculated in windows where $m_1$ is fixed (say, to $+1$). Within such a window, $I_h = J_1$ and $\langle m_1 m_2 \rangle$ can be simplified to $~ \langle m_2 \rangle$. Following Eq.~(\ref{steady}), $m_h$, in this case, behaves as a biased binary random telegraph noise (RTN) according to the following equation: 
\begin{equation}
    m_h \approx \operatorname{sign}\\ (J_1 \beta + \eta_h(t)\bigr).
\end{equation}
Here $\eta_h(t)$ is characterized by a correlation timescale $\tau_r^h$ which in our present analysis is $\lessapprox \tau_{\mathrm{s}}$. In this case, the switching rates from $\lambda_{+1 \to -1}$ and $\lambda_{-1 \to +1}$ are modulated from their unbiased values ($= 1/2\tau_r^h$) as follows,

\begin{equation}
\begin{aligned}
    \lambda_{+1 \to -1} &= \frac{1}{2\tau_r^h} \left(1 - J_1 \beta  \right) \\
     \lambda_{-1 \to +1} &= \frac{1}{2\tau_r^h} \left(1 + J_1 \beta  \right).
\end{aligned}
\end{equation}
This results in modulating the probability of $m_h$ dwelling in states $+1$ and $-1$ as follows,
\begin{equation}
    \begin{aligned}
        p_+ &= \frac{\lambda_{-1 \to +1}}{\lambda_{-1 \to +1} +  \lambda_{+1 \to -1}} = \frac{1 + J_1 \beta}{2} \\
         p_- &= \frac{\lambda_{+1 \to -1}}{\lambda_{-1 \to +1} +  \lambda_{+1 \to -1}} = \frac{1 - J_1 \beta}{2}
    \end{aligned}
\end{equation}
The time-average of $m_h$ is thus $\langle m_h \rangle = \left(p_+ - p_-\right) = J_1 \beta$. In the $\tau_r^h/\tau_{\mathrm{s}} \lessapprox 1$ limit, the autocorrelation function of $m_h$ will then be given as,

\begin{equation}
    R_{h}(\tau) \;=\; \langle 
    m_h(t)\,m_h(t+\tau) \rangle
\;=\; \,\exp \\ \Bigl(-\frac{\lvert \tau \rvert}{\tau_{r}^h}\Bigr).
\end{equation}

The Fourier Transform of this autocorrelation function gives the Power Spectral Density (PSD) $S_{h}(f)$ in the frequency domain as,

\begin{equation}
\begin{aligned}
 S_{h}(f) = \mathcal{F}\left[R_h(\tau)\right] &=  \int_{-\infty}^{+\infty}R_{h}(\tau) e^{-i2 \pi f \tau} d\tau \\
 &= \frac{2 \tau_r^h}{1 + (2\pi f \tau_r^h)^2}
\end{aligned}
\end{equation}

For $\tau_r^h/\tau_{\mathrm{s}} \lessapprox 1$, due to rapid updates of the hidden p-bit, $m_2$ can not instantaneously track $m_h$. Instead, it is then integrated by the RC-like low pass filter (with a time constant $\tau_{\mathrm{s}}$ and DC gain $J_2$) whose transfer function $H(f)$ is given by,
\begin{equation}
\begin{aligned}
H(f) &= \frac{J_2}{1 + i 2 \pi f \tau_{\mathrm{s}}} 
\end{aligned}
\end{equation}
Note that the synaptic filter passes $\langle m_h \rangle$ with a DC gain of $J_2$ that results in $I_2$ that has a mean $\mu = J_2 J_1 \beta$. However, in this limit of a fast hidden p-bit, the filtered $I_2$ can be approximated as Gaussian variable with standard deviation $\sigma$:
\begin{align}
    \sigma &= \sqrt{\int_{-\infty}^{+\infty} S_{h}(f) \mid H(f) \mid^2 df}  = \frac{J_2}{\sqrt{1 + \tau_{\mathrm{s}}/\tau_r^h}}.
 \label{eq:sigma_expression}
\end{align}
For $J_1 J_2 \beta^2 \gtrapprox  1$, we can neglect the local noise $\eta_2$ in the second computational p-bit yielding $m_2 = \operatorname{sign}(\beta I_2)$. Hence, $\langle m_2 \rangle$ can be evaluated and we could arrive at Eq.~(\ref{eq:mimijformula2}) of the main text:

\begin{align}
    \langle m_1 m_2 \rangle  = \operatorname{erf}\left(\frac{\mu}{\sqrt{2}\sigma}\right) = \operatorname{erf} \left(\frac{J_1 \beta}{\sqrt{2}} \sqrt{1 +  \frac{\tau_{\mathrm{s}}}{\tau_r^h}}\right),
    \label{eq: m1m2for3bitmodel}
\end{align}

\noindent where error function $\operatorname{erf}(x)$ is defined as $\operatorname{erf}(x) = \frac{2}{\sqrt{\pi}} \int_0^x e^{-t^2}\, dt$. 

Note that Eq. (\ref{eq: m1m2for3bitmodel}) is strictly valid in the limit when $\tau_r/\tau_{\mathrm{s}} \gg 1$ for the computational p-bits, $\tau_r^h/\tau_{\mathrm{s}} \lessapprox 1$ for the hidden p-bit, $J_1 \beta \ll 1$ and $J_2 J_1 \beta^2 \gtrapprox  1$. We notice that $\mu/\sigma$ facilitates the tunability in $\langle m_1 m_2 \rangle$ in this $\tau_r^h/\tau_{\mathrm{s}} \lessapprox 1$ regime. For larger $\mu/\sigma$ realized when the hidden p-bits are ultra-fast, the computational p-bits become strongly correlated. With a decrease in $\mu/\sigma$, $\langle m_1 m_2 \rangle$ too, decreases accordingly. \\

\begin{figure}[t]
    \centering \includegraphics[width=\columnwidth]{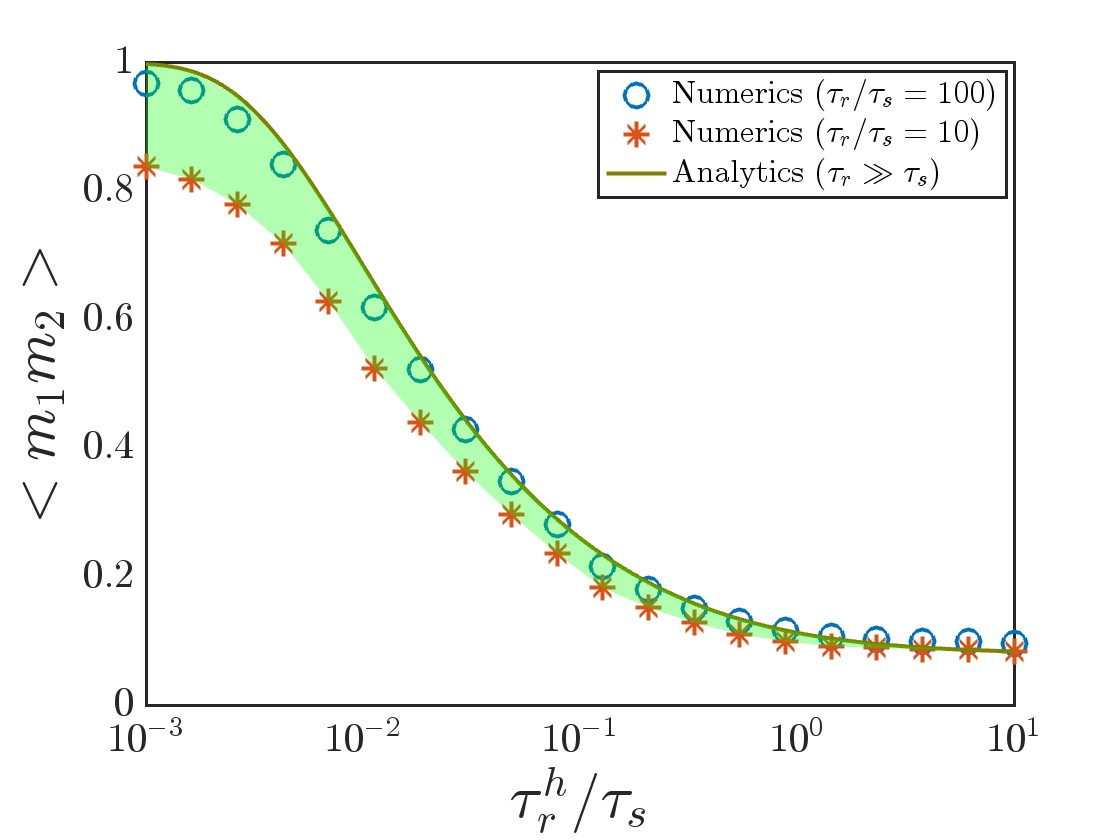}
    \caption{Effect of  $\tau_r/\tau_{\mathrm{s}}$ for the computational p-bits on the variation in $\langle m_1 m_2 \rangle$ as a function of $\tau_r^h/\tau_{\mathrm{s}}$. The analytics (olive green line), derived under the assumption $\tau_r/ \tau_{\mathrm{s}} \gg 1$, shows better agreement to the numerical model for $\tau_r/\tau_{\mathrm{s}} = 100$ (blue) compared to $\tau_r/\tau_{\mathrm{s}} = 10$ (orange). The green shaded region further illustrates that larger $\tau_r/\tau_{\mathrm{s}}$ lead to a closer agreement to the analytics.}
    \label{fig:1_supp}
\end{figure}

\section{Hidden p-bit model: $\tau_r^{1(2)}$ dependence}

A crucial assumption connecting the numerical results presented in the main text and analytical derivations in appendix~\ref{supp: 3pbitmodel} is faster synaptic response times w.r.t correlation timescales $\tau_r^{1(2)}$ for the computational p-bits, i.e., $\tau_r^1/\tau_{\mathrm{s}} \gg 1$ and $\tau_r^2/\tau_{\mathrm{s}} \gg 1$. This is a key requirement to enable tunability in correlations $\langle m_1 m_2 \rangle$ via the hidden p-bit. Indeed, numerical simulations of the hidden p-bit model presented in Fig.~\ref{fig:1_supp} agree closely with the analytical model more for $\tau_r/\tau_{\mathrm{s}} = 100$ than $\tau_r/\tau_{\mathrm{s}}=10$. \\

\section{Time-Averaged Output in a p-bit}
\label{supp: Effective2pbitmodel}
In this section, we evaluate the time-averaged output of a p-bit for a \textit{fixed} input. From Eq.~\eqref{steady}, if the input to the p-bit is $I$, the output $m$ is given by
\begin{equation}
    m = \operatorname{sign}\left[\tanh(\beta I) + \eta\right],
\end{equation}
where $\eta$ is a uniformly distributed random variable in $[-1, 1]$. The probability that $m = \pm 1$ is then,
\begin{equation}
    p(m = \pm 1) = \frac{1}{2}\left(1 \pm \tanh(\beta I)\right).
\end{equation}

\noindent In that case, the time-averaged output (correlator) $\langle m \rangle$ can be computed as
\begin{equation}
    \langle m \rangle = p\cdot(+1) + p\cdot(-1) = \tanh(\beta I).
    \label{eq:mimj2pbitmodel}
\end{equation}




\bibliographystyle{apsrev4-2}
\bibliography{apssamp}

\end{document}